\documentclass[doublecol]{epl2} 

\usepackage{amsmath}
\usepackage{amssymb}
\usepackage{gensymb}
\usepackage{graphicx,bm,units,yfonts}
\usepackage[table]{xcolor}
\usepackage{hyperref}
\usepackage{color}
\usepackage{lipsum}

\usepackage{soul}

\newcommand{\CM}{{\mathbb C}}

\newcommand{\SM}{{\mathbb S}}
\newcommand{\TM}{{\mathbb T}}
\newcommand{\ZM}{{\mathbb Z}}

\title{Pumping with symmetry}

\author{J. A. Iglesias Mart\'inez\inst{1} \and M. Kadic\inst{1} \and V. Laude\inst{1} \and E. Prodan\inst{2}}
\shortauthor{J. A. Iglesias Mart\'inez \etal}

\institute{
 \inst{1} Université de Franche-Comté, CNRS, Institut FEMTO-ST, Besançon, France\\
 \inst{2} Department of Physics, Yeshiva University, New York, USA
 }

\abstract{Re-configurable materials and meta-materials can jump between space symmetry classes during their deformations. Here, we introduce the concept of singular symmetry enhancement, which refers to an abrupt jump to a higher symmetry class accompanied by an un-avoidable reduction in the number of dispersion bands of the excitations of the material. Such phenomenon prompts closings of some of the spectral resonant gaps along singular manifolds in a parameter space. In this work, we demonstrate that these singular manifolds {can} carry topological charges. As a concrete example, we show that a deformation of an acoustic crystal that encircles a $p11g$-symmetric configuration of {an array of} cavity resonators results in an adiabatic cycle that carries a Chern number in the bulk and displays Thouless pumping at the edges. {This points to a very general guiding principle for recognizing cyclic adiabatic processes with high potential for topological pumping in complex materials and meta-materials, which rests entirely on symmetry arguments}. }
\begin{document}

\maketitle

It has been recently recognized that the space symmetry of materials can be a rich source of topological effects. For example, in topological quantum chemistry, the materials are divided in homotopy classes such that two systems from two different classes cannot be continuously deformed into each other without closing a bulk spectral gap or breaking the space symmetries associated with the classes \cite{PoNatComm2017,BradlynNature2017,CanoPRL2018,VergnioryPRE2017}. It has also been recognized that robust topological bulk-boundary correspondences can be induced by space symmetries under adiabatic pumping conditions \cite{PeriScience2020}. Space symmetries also play a central role in the topological effects observed in higher order topological insulators \cite{SchindlerSA2018} and in other manifestations of bulk-boundary correspondences, such as in topological corner modes \cite{BenalcazarScience2027} and topological screw dislocations \cite{LinNatMat2022}.

Our present work opens another perspective on space symmetries, specifically, on cyclic deformations of materials in the presence of symmetries. As it is well known \cite{NiuRMP2010}, an adiabatic cycle can generate non-trivial topology in the bulk of a material and a topological spectral flow at its boundaries. The prototypical source of such phenomena is the Rice-Mele model \cite{RiceMele1982} of polyacetylene $[{\rm C}_2{\rm H}_2]_n$. It has two parameters that quantify the fluctuations of the hopping coefficients  and an ensuing staggered potential under possible dimerizations of the ideal chain. The model displays a gapless energy spectrum at a singular point, where the two parameters are zero, and a gapped spectrum otherwise. A closed loop encircling this singular point of the parameter space supports a non-trivial Chern number \cite{NiuRMP2010}. This is often invoked as an example where the symmetry is irrelevant and where the principles at work are entirely topological. We argue here that this is a very narrow point of view, which misses a wider picture that can be very revealing when it comes to identifying materials that support similar topological effects.

\begin{figure}[t]
\center
  \includegraphics[width=0.6\linewidth]{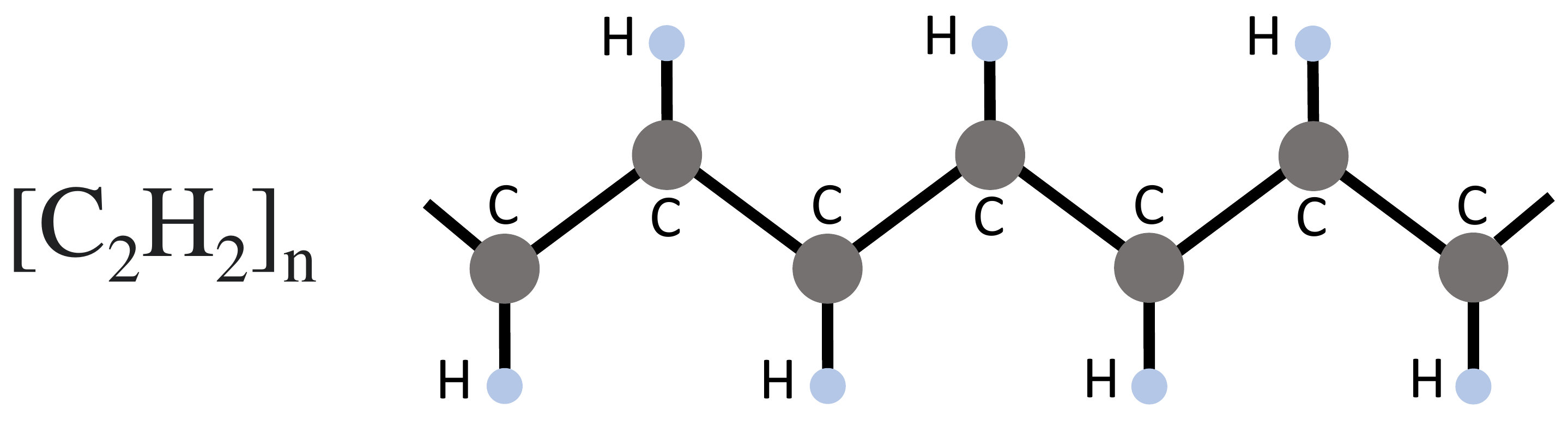}
  \caption{Ball-and-stick model of un-dimerized polyacetylene.}
 \label{Fig:Poly}
\end{figure}

To open the discussion, we point out the glide-reflection symmetry of the un-dimerized polyacetylene chain, which is quite evident from its structure reproduced in fig.~\ref{Fig:Poly}. The singular point in the parameter space of the Rice-Mele model, which carries the topological charge, exists precisely because of this symmetry.
Away from this singular point, the glide-reflection symmetry of the polyacetylene chain is removed, so that only discrete translations remain ($p1$ symmetry) and the energy spectrum is gapped. Thus, we are dealing with a parameter space with predominantly $p1$ symmetry and with one singular point where the symmetry is enhanced. The main point we want to communicate is that, even without a tight-binding model for a material or metamaterial, we can still identify {cyclic adiabatic processes with potential for topological pumping,} based entirely on symmetry principles. Indeed, we will show in this work that the topological adiabatic cycle in polyacetylene is not an isolated occurrence and that, in fact, topological cycles are prone to occur and very easy to identify around the sub-manifolds of the parameter space that carry enhanced symmetries.

\begin{figure}[t]
\center
  \includegraphics[width=\linewidth]{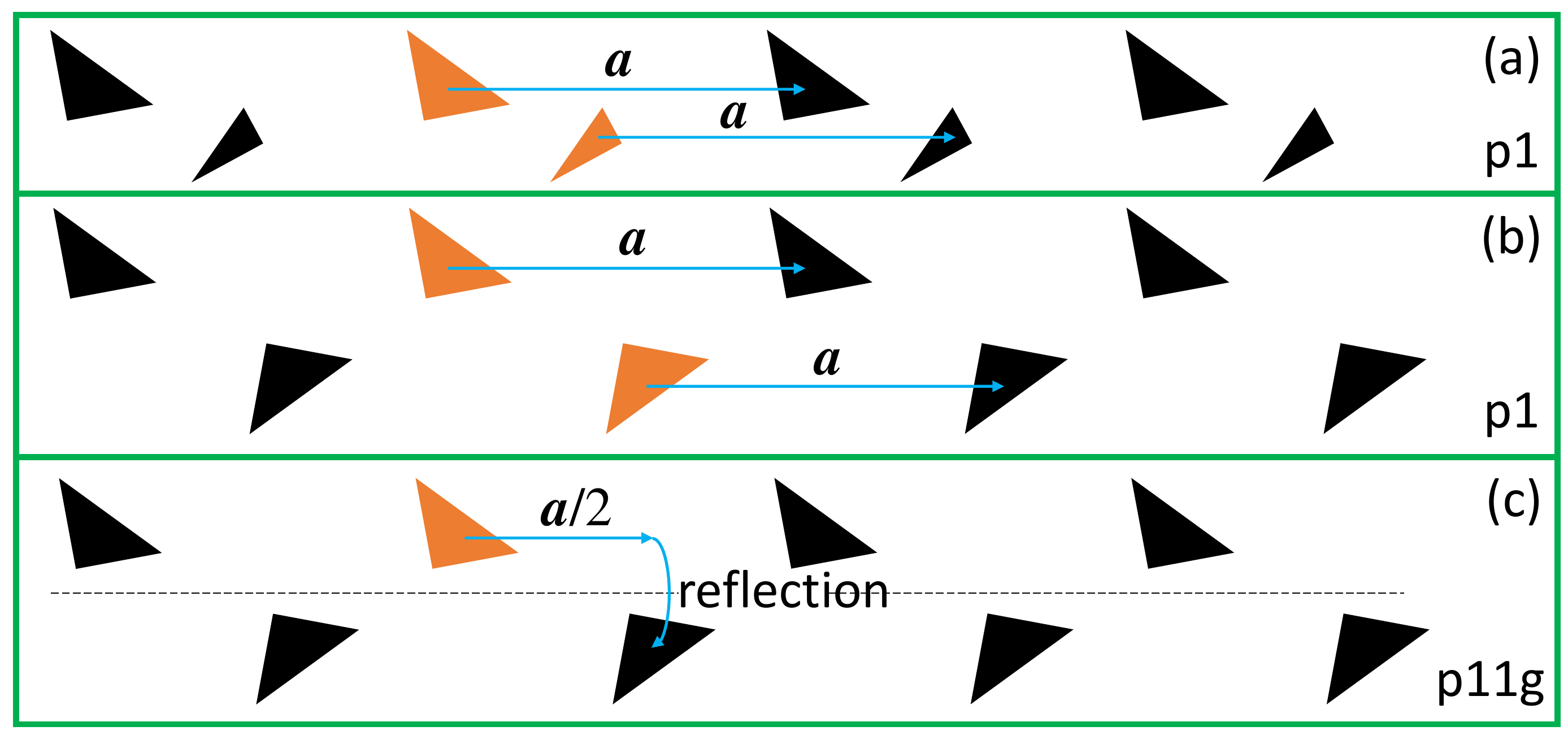}
  \caption{(a) Symmetric pattern generated by acting with $p1$-group on the two seeding shapes shown in color{, i.e. using only discrete lattice translations.}
  {$a$ is the lattice constant.}
  {(b) Same as (a), but with the seeding shapes being glide-reflection images of each other.}
  (c) Pattern (b) is reproduced by acting with $p11g$-group on a single seeding shape{, i.e. by repeatedly applying a half-shift followed by a glide-reflection.}}
 \label{Fig:SP}
\end{figure}

\begin{figure}[t] 
\center
  \includegraphics[width=\linewidth]{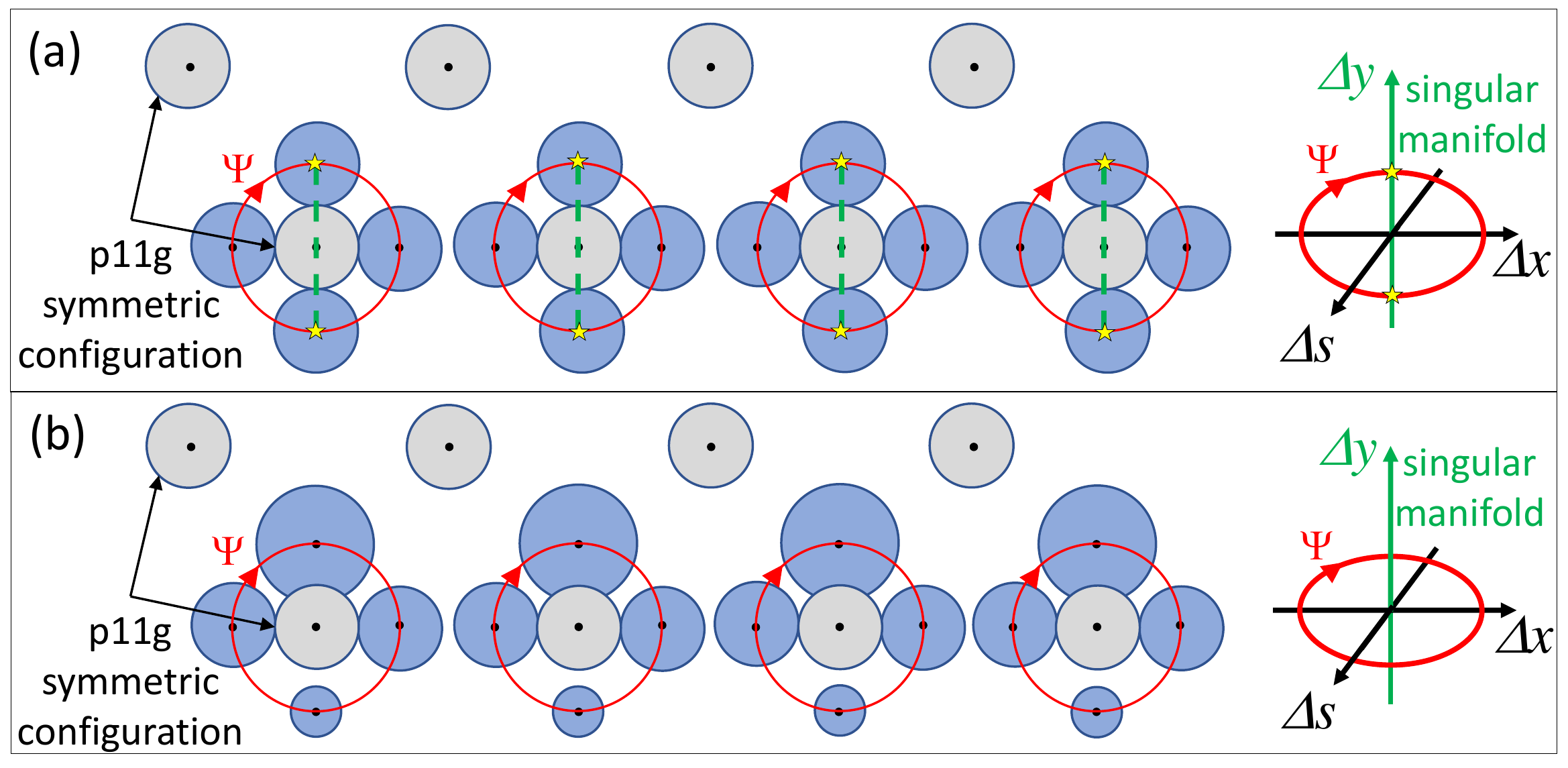}
  \caption{{Adiabatic deformations of the materials described in this work can be generated by modifications of position (hence $x,y$ coordinates) and size (hence scaling by a parameter $s$) of a seed resonator. When going around the reference $p11g$-symmetric configuration, shown in gray and corresponding to the origin of the parameter space $\Delta x=\Delta y=\Delta s=0$}: (a) {An a}diabatic cycle generated only with displacements inherently intersects the singular manifold shown in green; (b) {whereas an a}diabatic cycle generated with displacements plus scalings  can encircle the critical manifold. The shapes in these diagrams represent self-coupling resonators. In blue, we show a continuous sequence of configurations indexed by the adiabatic parameter $\Psi$.
  The diagrams on the right depict the corresponding loops in parameter space.
  {Configurations drawn with starred yellow centers are $p11g$-symmetric. Note that, in (b), the adiabatic loop goes behind at the top and in front at the bottom of the singular manifold, hence avoiding these $p11g$-symmetric configurations.}}
 \label{Fig:AC1}
\end{figure}

To formulate the latter more precisely, let us place the discussion in the context of the seven frieze groups, which are the discrete space-groups of planar strips \cite{ConwayBook}, hence appropriate for the investigation of periodic quasi 1-dimensional physical systems. We also switch from molecular systems to meta-materials, for which we generate symmetric patterns where the positions, orientations and shapes of the resonators all influence the dynamics of the collective resonant modes \cite{LuxArxiv2022}. Specifically, starting from one or more seeding shapes, later regarded as seeding resonators, we apply all plane transformations contained in a particular frieze group and generate patterns displaying a desired symmetry. In regards to the complexity of their symmetries, at one end stands the $p1$ group, which includes only the discrete translations of the primitive cell. At the opposite end stands the $p2mm$ group, which incorporates the maximal set of allowed discrete symmetries, that is, horizontal translations and horizontal/vertical reflections. Regardless of their complexity, all symmetric patterns can be generated with the $p1$ group, when the latter acts on an appropriate set of seeding resonators forming the primitive cell, as illustrated in fig.~\ref{Fig:SP}(a). However, if the seeding resonators have particular shapes, locations and orientations as in fig.~\ref{Fig:SP}(b), the symmetry of the pattern can be enhanced to other frieze groups and a smaller set of seeding resonators is needed, as illustrated in fig.~\ref{Fig:SP}(c) for the case of the glide-reflection symmetry (frieze group $p11g$). Under the slightest generic perturbation of the position, shape or orientation of resonators, the pattern falls back to the $p1$ symmetry. The point we want to make is that the patterns with symmetries other than $p1$ form isolated manifolds in the space of symmetric patterns, and these manifolds are surrounded and connected by the space of $p1$ symmetric patterns. We demonstrate here that some of these isolated manifolds carry topological charges. 

A decisive factor that must be taken into account is the number of distinct energy bands that can be produced with a given number of seeding resonators. If each resonator carries one resonant mode, then the $p1$-symmetric pattern seen in fig.~\ref{Fig:SP}(a) produces two gapped resonant energy bands, generically. In contrast, the $p11g$-symmetric pattern seen in fig.~\ref{Fig:SP}(c) can produce only a single resonant energy band, regardless of the couplings \cite{IglesiasPRB2022}. Thus, the spectrum will be ungapped along a manifold of parameters carrying the $p11g$-symmetry. However, not every enhancement/reduction of symmetry leads to the phenomena advertised here. If we start with four arbitrary seeding resonators, we can generate a $p1$-symmetric pattern with four resonators in the primitive cell. By continuously changing the shape, orientation and locations of the seeding resonators, we can achieve the $p2mm$ symmetry, in which case the pattern can be generated from a single seeding resonator. Yet, both cases can display four separated energy bands, a counting that is based on the K-theories of these space groups (see {\it e.g.} fig.~7.3 in \cite{LuxArxiv2022}). In this case, the symmetry enhancement does not display a singular character. In contradistinction, the enhancement from frieze group  $p1m1$ (vertical reflection) to $p2mg$ (vertical reflection and glide-reflection) does, because the analysis here is very similar to the one for $p1 \to p11g$ enhancement: $p1m1$ and $p2mg$ belong to the same isomorphism class, hence they have identical $K$-theories, but the number of needed seeding resonators drops by one for $p2mg$. Hence, the number of energy bands that can be produced by a  $p2mg$-symmetric pattern is necessarily {\it lower} than that produced by a $p1m1$-symmetric one. While a more thorough analysis will be reported elsewhere, we can already state the general {guiding} principle at work here, namely, the reduction in the number of energy bands that can be produced with a given pattern when the symmetry enhancement occurs.

\begin{figure}[t] 
\center
  \includegraphics[width=\linewidth]{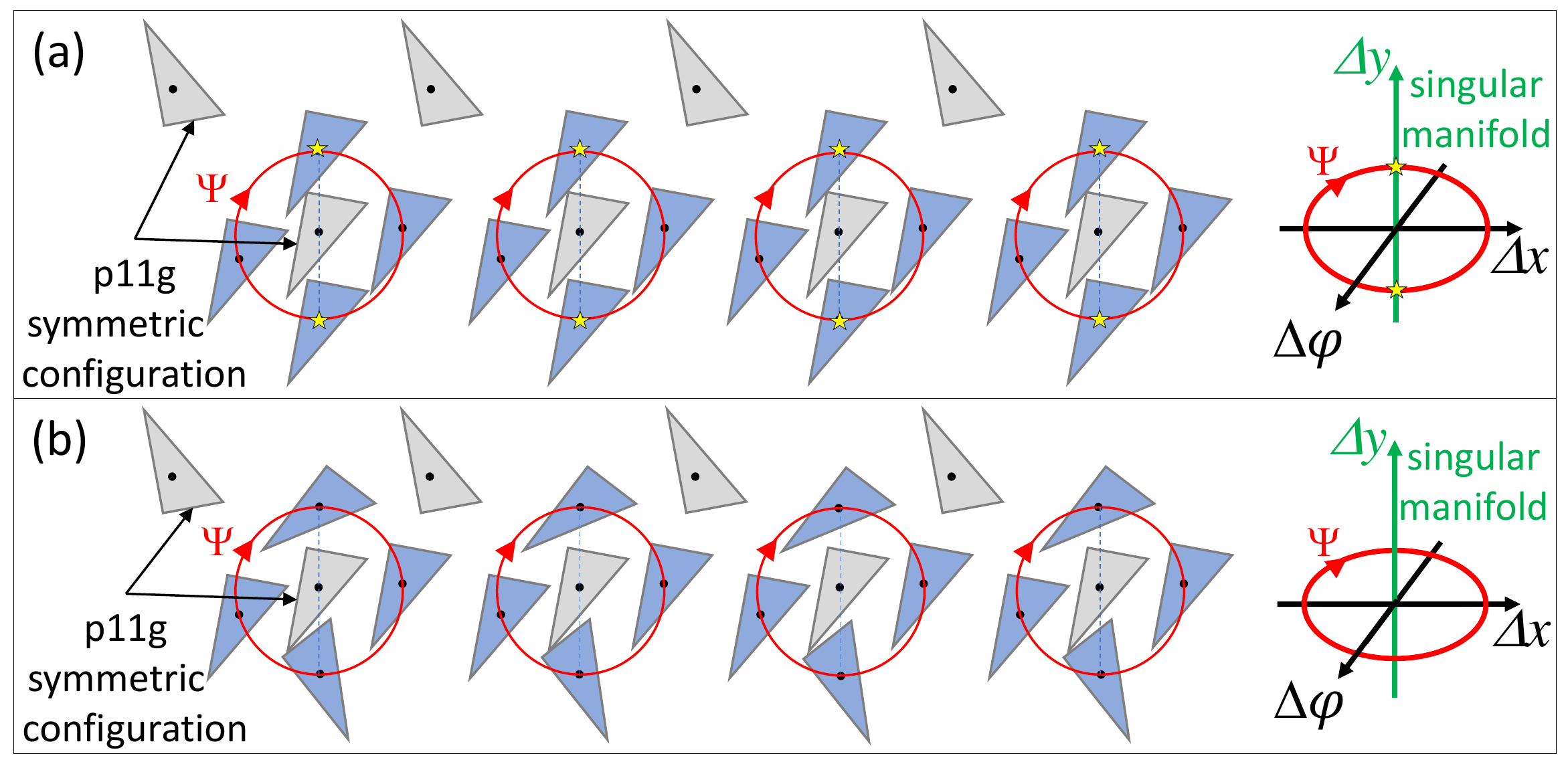}
  \caption{Same as fig.~\ref{Fig:AC1} but with scalings replaced by rotations.
  (a) {An adiabatic cycle generated only with displacements inherently intersects the singular manifold shown in green; (b) whereas an adiabatic cycle generated with displacements plus rotations (by parameter $\Delta \varphi$) can encircle the critical manifold.
  The shapes in these diagrams represent self-coupling resonators.
  The configuration shown in gray is the $p11g$-symmetric configuration at the origin of parameter space, or $\Delta x=\Delta y=\Delta \varphi=0$.
  In blue, we show a continuous sequence of configurations indexed by the adiabatic parameter $\Psi$.
  The diagrams on the right depict the corresponding loops in parameter space.
  Configurations drawn with starred yellow centers are $p11g$-symmetric.}}
 \label{Fig:AC2}
\end{figure}

Once we identified a symmetry enhancement that leads to a singular manifold in the space of parameters, the next step is to construct adiabatic cycles that encircle this manifold. They can {\it all} be obtained by deforming the seeding resonators. In general, a seeding resonator has an infinite dimensional configuration space{, so there are many opportunities to engineer deformation spaces of different topologies and dimensions. In the present study, however, we} restrict ourselves to lower dimensional deformation spaces, by only allowing specific {actions on} the resonators. In fig.~\ref{Fig:AC1}, for example, we consider a pair of spherical seeding resonators, the first one with constant radius $r_0$ and the second one initially {of same radius $r_0$} and fixed to the glide-reflection symmetric position {(see the gray configuration in fig.~\ref{Fig:AC1}). From this $p11g$-symmetric reference configuration, we allow displacements and scalings (by parameter $s$) of the second seeding resonator. Thus,} the configuration space $({\Delta} x,{\Delta} y,{\Delta}s)$ is 3-dimensional. We then see that the space of $p11g$-symmetric patterns is 1-dimensional and is represented by $(0,{\Delta}y,{0})$, or the vertical dashed line in fig.~\ref{Fig:AC1}(a). Any closed loop $({\Delta}x(\Psi),{\Delta}y(\Psi),{0})$ around the {reference} point {$(0,0,{0})$}, composed of displacements only, will intersect the singular manifold at least twice. As shown in fig.~\ref{Fig:AC1}(b), however, we can encircle the $p11g$-symmetric phase by using displacements and scalings with a closed loop $({\Delta} x(\Psi), {\Delta}y(\Psi),{\Delta}s(\Psi))$ that never crosses the singular manifold {and actually encircles it}.
{Note that parameter space is represented in fig.~\ref{Fig:AC1} as $(\Delta x, \Delta y,{\Delta} s)$ so that the reference configuration $(\Delta x=0, \Delta y=0,{\Delta}s=0$ sits at the origin.}

Another example is shown in fig.~\ref{Fig:AC2}, where the shapes and sizes of the seeding resonators are matched and fixed, but we allow displacements and rotations $\Delta \varphi$ of the second seeding resonator. This produces again a closed loop $({\Delta}x(\Psi),{\Delta}y(\Psi), \Delta \varphi(\Psi))$ in the 3-dimensional configuration space that encircles the critical manifold without crossing it.
{Again, the origin of parameter space represents the $p11g$-symmetric reference phase.}

\begin{figure}[t] 
\center
  \includegraphics[width=0.99\linewidth]{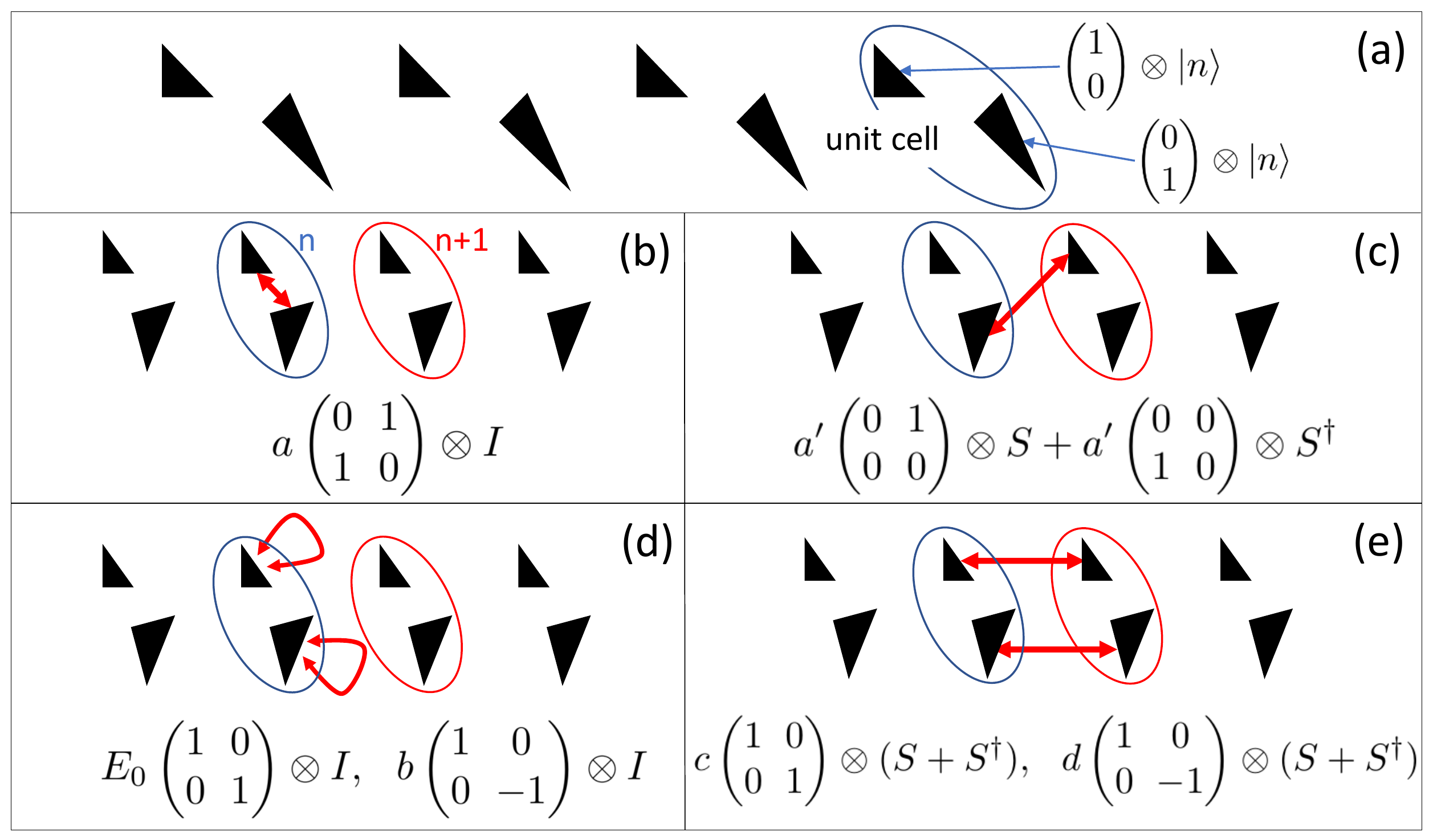}
  \caption{(a) The vectors of the Hilbert space $\CM^2 \otimes \ell^2(\ZM)$ corresponding to the resonant modes carried by the two resonators; (b-e) The dominant coupling matrices for a generic $p1$-symmetric pattern with two resonators per primitive cell.
  {$a$ and $a'$ are intra- and inter-cell cross-couplings, respectively.
  $E_0 \pm b$ are the on-site self energies.
  $c \pm d$ are inter-cell couplings for each site.}
  }
 \label{Fig:Coup}
\end{figure}

The big claim of our work is that, {at least in the case when the seed resonators carry a single mode each}, the topologically non-trivial cycles {described above} automatically translate into Thouless pumpings provided the system displays a gap in its resonant spectrum, and {\it no} tight-binding models are needed to understand this phenomenon. {Indeed, in the proposed scenario, the coupling matrices between the resonators are functions of only the three specified parameters, but their functional dependence can be arbitrary. Now, during an adiabatic cycle, the collective resonant states can be resolved over the adiabatic variable $\Psi$ and quasi-momentum $k$, which both live on circles. Thus, the collective resonant states with frequencies below the spectral gap supply a vector bundle over the 2-torus. Such bundles are generated by acting with a $2 \times 2$ projection matrix $P(\Psi,k)$ on the fixed space $\CM^2$ (see {\it e.g.} Eq.~\eqref{Eq:Proj} for an explicit expression). Every projection obeys the constraints $P=P^\dagger=P^2$ and, as a result, they take the form 
\begin{equation}\label{Eq:GenProj}
P = \begin{pmatrix} \alpha & \sqrt{\alpha(1-\alpha)} e^{i \phi} \\
\sqrt{\alpha(1-\alpha)} e^{-i \phi}  & 1 -\alpha \end{pmatrix}
\end{equation}
where $\alpha$ is a real parameter from the interval $[0,1]$ and $\phi$ is also a real parameter from the interval $[0,2 \pi)$. Key here is that, with our 3-dimensional deformation space and with the freedom to choose the functional dependence of the coupling matrices, we can sample any desired projection, which requires only two parameters, as seen in Eq.~\eqref{Eq:GenProj}. Now, consider one configuration with gapped spectrum, which can be always expanded into a small adiabatic loop of configurations without closing the gap. The resulting bundle over the 2-torus is obviously trivial, hence it carries a Chern number zero. Consider now an adiabatic loop that encircles the singular manifold carrying the $p11g$ symmetry. We claim that the resulted bundle over the 2-torus is topologically distinct from first bundle we previously constructed. Indeed, if we can modify the functional dependence of the couplings between the resonators such that the second bundle is deformed into the first one, then we should be able to contract the second adiabatic loop to a point, without closing the gap. But this is impossible because, in the process, we will necessarily touch the singular manifold where there is only one spectral band in the spectrum. Thus, the second bundle must carry a nontrivial Chern number. 
}

{Once we established that the vector bundle supported by the loop encircling the $p11g$-singular manifold carries a non-trivial Chern number, given the robustness of the latter, we can consider additional deformations of the resonators as well as turning on additional degrees of freedom. As long as the spectral gap remains open for the entire adiabatic cycle, none of the above actions can destroy the topological character of the loop. This conclusion is definitely aided by the particular setting we started with. If the resonators would have carried more than one degree of freedom, then it is very likely that some of the bands or composite bands will not carry a Chern number under the proposed scheme. Therefore, we want to be clear that we are not announcing a theorem here, but rather a guiding principle for detecting cyclic adiabatic processes with high potential for implementing topological pumps. In general, a check will still be necessary to confirm the sought properties.  
}

\begin{figure}[t] 
\center
  \includegraphics[width=0.99\linewidth]{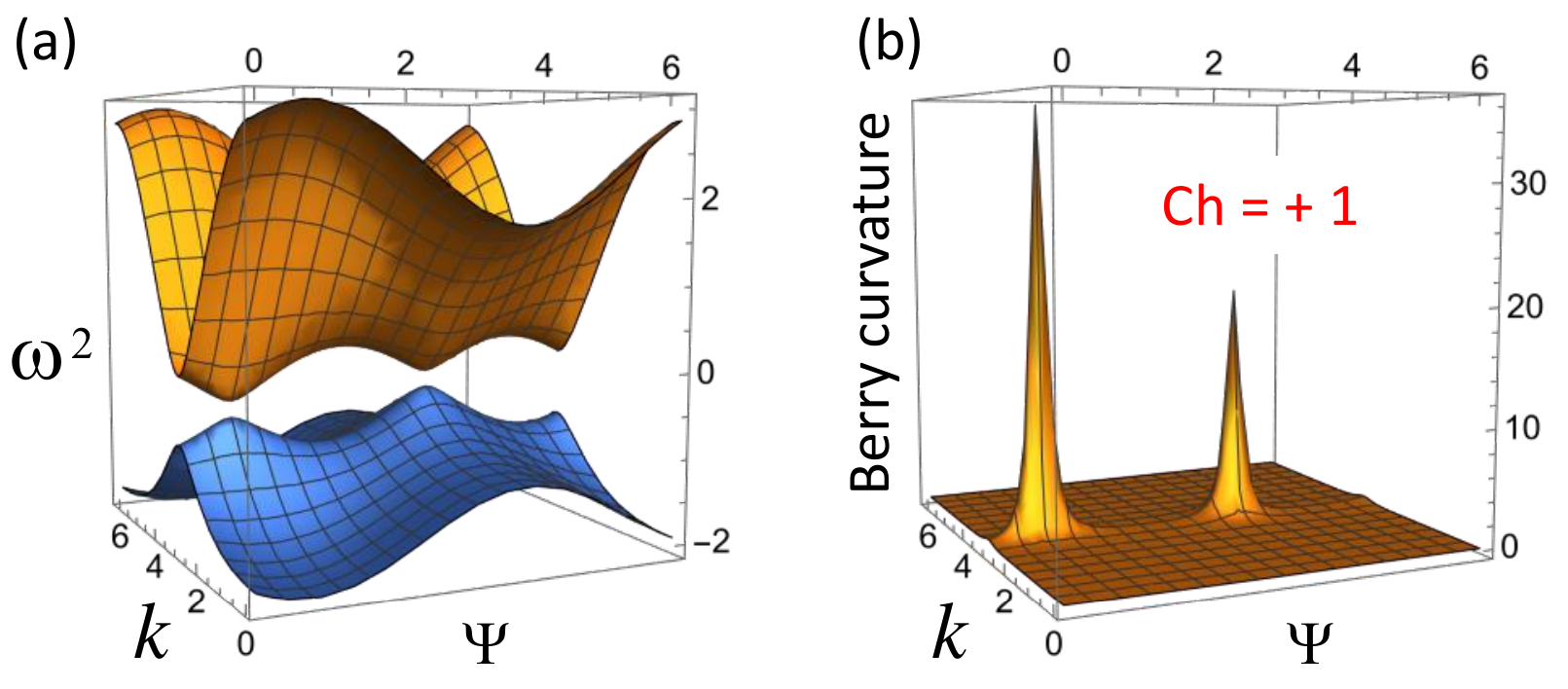}
  \caption{Resonant spectrum (a) and Berry curvature (b) for the loop $a = 1.0 + 0.4\cos(\Psi)$, $a' = 1.0 + 0.4 \sin(\Psi)$, $b = 0$, $c = 0.4 + 0.1\cos(\Psi - \pi/4)$ and $d = - 0.1\cos(\Psi - \pi/4)$. The expressions of parameters are consistent with the loop in fig.~\ref{Fig:AC2}, where $b=0$ because the resonators are identical and $c+d$ is a constant because the top row is not modified. The $\pi/4$ phase is an arbitrary choice that makes no difference.}
 \label{Fig:Chern}
\end{figure}

{We will confirm the} topological nature of the cycle from fig.~\ref{Fig:AC2} using both tight-binding and continuum media simulations. The pattern of resonators discussed so far have two modes per repeating cell, hence the Hilbert space of the resonating modes is spanned by the vectors $\xi \otimes |n\rangle$, with $\xi$ a column vector with two complex entries and $n \in \ZM$. This particular Hilbert space is usually denoted as $\CM^2 \otimes \ell^2(\ZM)$. The precise distribution of these vectors on the resonators is shown in fig.~\ref{Fig:Coup}(a). In figs.~\ref{Fig:Coup}(b-e), we list the nearest-neighbor couplings and their contributions to the dynamical matrix. According to those couplings, the dynamical matrix $H$ which determines the resonant pulsations $H|\psi\rangle = \omega^2 |\psi \rangle$ of the resonator pattern can be approximated as
\begin{equation*}
\begin{aligned}
H = & E_0 + a \sigma_1 \otimes I + a' \, (\sigma_- \otimes S + \sigma_+\otimes S^\dagger) + b \, \sigma_3 \otimes I \\
& + \tfrac{c}{2} \sigma_0 \otimes( S + S^\dagger) + \tfrac{d}{2} \sigma_3 \otimes ( S + S^\dagger),
\end{aligned}
\end{equation*}
where $S|n\rangle = |n+1\rangle$ is the shift operator, $E_0$ is the average resonant energy of the two seeding resonators, $\sigma_0$ is the $2 \times 2$ identity matrix and the other $\sigma$'s are Pauli's matrices.
{The terms included encompass intra-cell couplings as well as inter-cell couplings up to the nearest neighbor.}
The terms left out involve second and higher nearest neighbor couplings, which are relatively small. We pass to the momentum space using the substitution $S^{\pm 1} \to e^{\pm \imath k}$, to find two energy bands 
\begin{equation*}
E_\pm(k) = E_0 + c \cos(k) \pm \sqrt{(b+d\cos(k))^2 + |a+a'e^{\imath k}|^2}
\end{equation*} 
separated by a gap. When the symmetry of the pattern is enhanced to $p11g$, then necessarily $a = a'$ and $b=d=0$;
{$E_\pm(k) = E_0 + (c \pm 2a) \cos(k)$}
 and the band spectrum is gapless.
 We recall that this closing of the energy gap is un-avoidable and that it does not depend on the simplified Hamiltonian we used. Under an adiabatic cycle parameterized by $\Psi \in \SM^1$, the Bloch Hamiltonian depends on two parameters $(\Psi,k)$ that live on torus $\TM^2$, hence we can evaluate a Chern number as
$$
{\rm Ch}(P) = \int_{\TM^2}  d\Psi \, dk \, F(\Psi,k), 
$$
where $P$ is the spectral projector onto the lower dispersion band, which can be conveniently computed as
\begin{equation}\label{Eq:Proj}
P(\Psi,k) = \frac{H(\Psi,k) - E_+(\Psi,k) \, I_{2\times 2} }{E_{-}(\Psi,k) - E_{+}(\Psi,k)},
\end{equation}
and $F(\Psi,k)$ is the Berry curvature,
$$
F(\Psi,k) = \tfrac{\imath}{2\pi} \ {\rm Tr}\Big( P(\Psi,k) \big [\partial_\Psi P(\Psi,k),\partial_k P(\Psi,k)\big]\Big ).
$$
Note that parameter $c$ does not enter in any of the last three equations because its corresponding contribution to $H(\Psi,k)$ just shifts $E_{\pm}(\Psi,k)$ by the same amount. Thus, we are effectively dealing with a 4-dimensional parameter space and, as such, the critical manifold $a-a'=b=d=0$ has dimension 1. Now, with these tools, we can verify that any loop that encircles the 1-dimensional critical manifold carries a non-trivial Chern number. Due to the topological character of the statement, it is enough to check this statement for one loop and this is confirmed in fig.~\ref{Fig:Chern}, where a parametrization consistent with the process illustrated in fig.~\ref{Fig:AC2} is considered.

\begin{figure}[t] 
\center
  \includegraphics[width=0.99\linewidth]{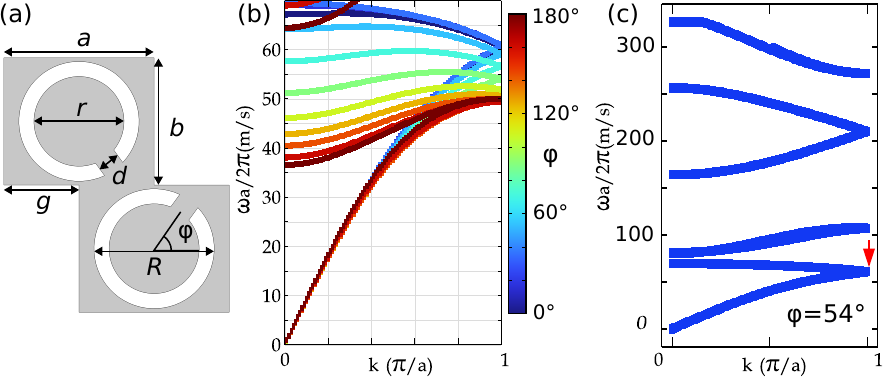}
  \caption{Tuning the resonating structure: (a) Unit cell of the $p11g$-symmetric configuration of a chain of C-shaped acoustic resonators. The gray region indicate the domain of the acoustic wave propagation. Hard-wall boundary conditions are applied at the top and bottom of the cell, as well as on the walls of the C-shape. In units of $a$, the values of the marked parameters are $b = 0.85 a$, $d = 0.15 a$, $r = 0.3 a$, and $R = 0.4 a$. The glide parameter $g$ is fixed at $a/2$. (b) Bulk dispersion as function of the resonator orientations. (c) Bulk dispersion of the tuned structure, with a mark showing the targeted band splitting. The simulations were generated with COMSOL MultiPhysics software.}
 \label{Fig:Tuning}
\end{figure}

\begin{figure}[t] 
\center
  \includegraphics[width=\linewidth]{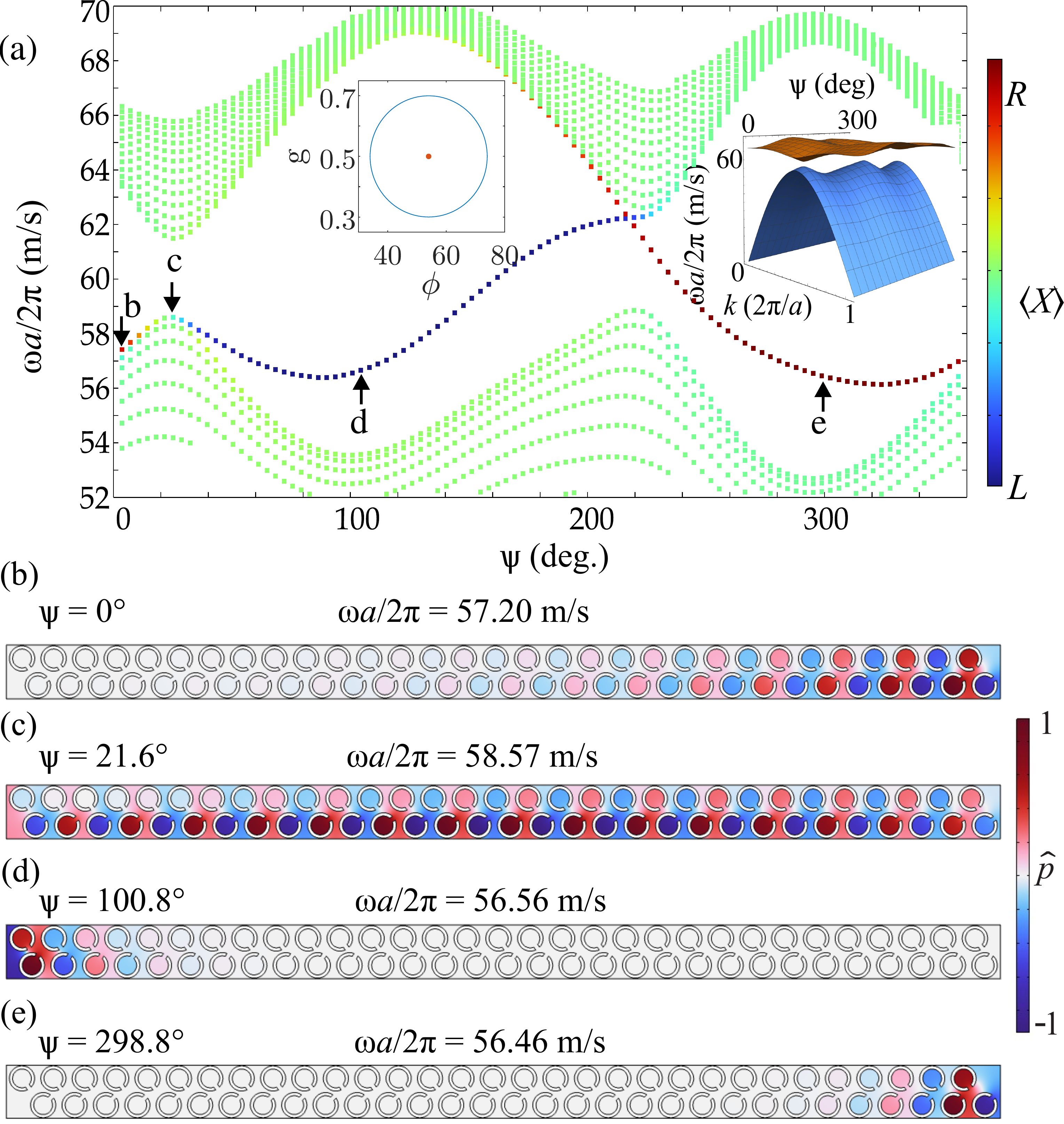}
  \caption{(a) Spectral flow of the resonant spectrum against the pumping parameter $\Psi$ for a finite acoustic crystal with 31 unit cells.
  The cycle parametrized by pumping parameter $\Psi$ encircles the glide symmetric point in parameter space $(g, \phi)$, as the left inset depicts. Topological edge modes are seen crossing the bulk gap of the dispersion diagram. Their localization, as measured by the center of mass $\langle X \rangle$ of the modes, is encoded in color.  (Right inset) Resonant spectrum of the infinite acoustic crystal (compare with fig. \ref{Fig:Chern}(a)). (b-e) Samples of spatial profiles of modes, reported as the pressure field $\hat p$, collected at the arrows indicated in panel (a).}
 \label{Fig:TP}
\end{figure}

\bigskip
We now demonstrate {how our guiding principle unfolds for} an actual acoustic crystal. Its building elements are the C-shaped Helmholtz resonators with the unit cell shown in fig.~\ref{Fig:Tuning}(a) {and full crystal shown in fig.~\ref{Fig:TP}(b-e)}. The out of page dimension of the resonators is small such that the low frequency resonant modes are all uniform in the direction perpendicular to the page. {We hope the reader will agree with us that, a priori, it is not clear what cyclic actions on such acoustic crystal will result in Thouless pumping.}  {According to our guidelines, the} first task is to optimize its $p11g$-symmetric configuration such that a breaking of the $p11g$-symmetry produces a complete gap in the dispersion spectrum. In fig.~\ref{Fig:Tuning}(b), we show the evolution of the dispersion bands as a function of the orientation of resonators in $p11g$-symmetric configurations and, from that data, we chose the angle $\varphi=54^\circ$ giving the dispersion shown in fig.~\ref{Fig:Tuning}(c). The touching of the dispersion bands marked there is protected by the $p11g$-symmetry and the loss of this protection will open a local spectral gap. The particular geometry of the dispersion bands then assures us that this local band splitting develops into a complete spectral gap. 

\begin{figure}[t] 
\center
  \includegraphics[width=\linewidth]{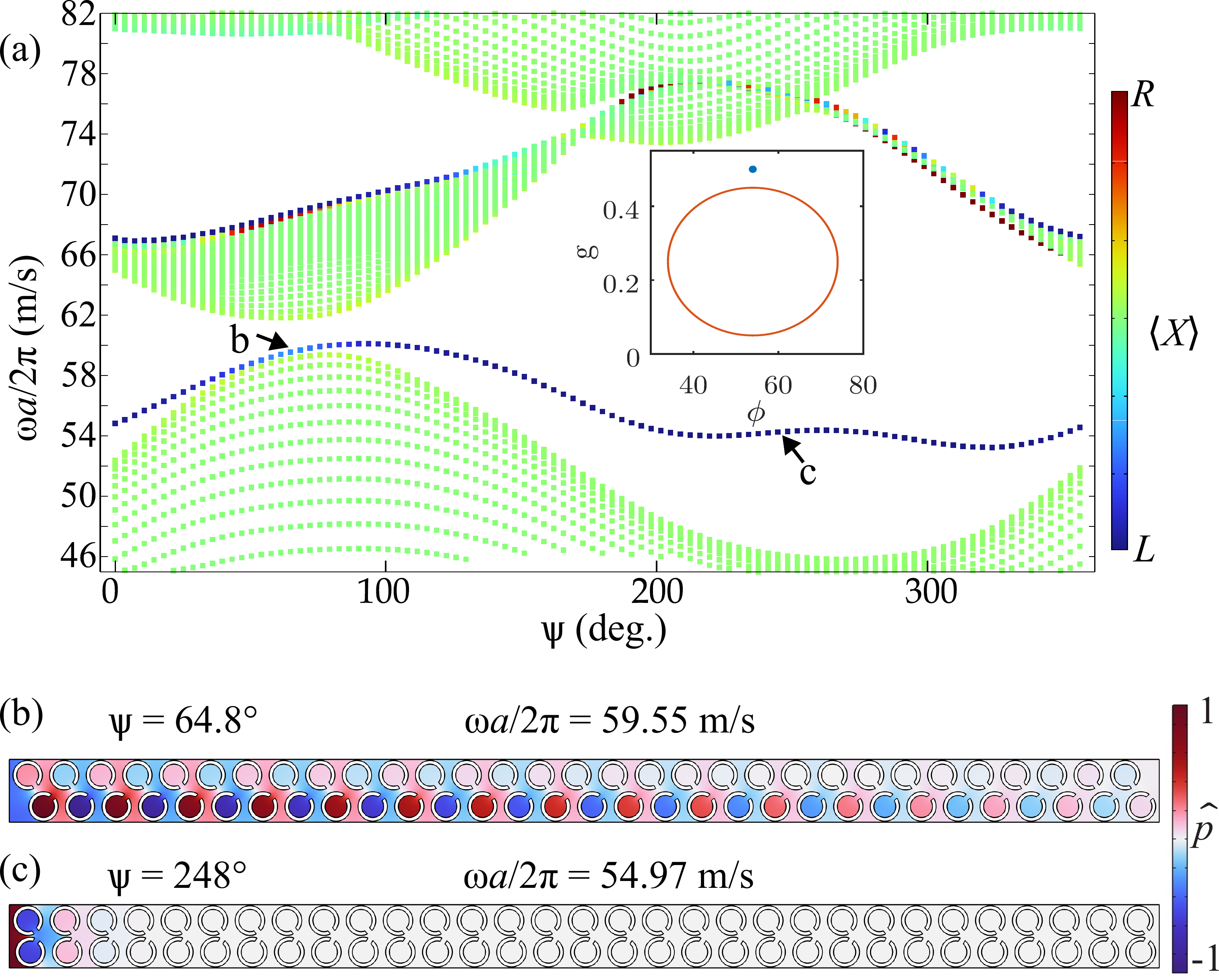}
  \caption{(a) Spectral flow of the resonant spectrum against the pumping parameter $\Psi$ for a finite acoustic crystal with 31 unit cells.
  The cycle parametrized by pumping parameter $\Psi$ does not encircle the glide symmetric point in parameter space $(g, \phi)$, as the inset depicts. The localization of topological edge modes, as measured by their center of mass $\langle X \rangle$, is encoded in color. (b-c) Samples of spatial profiles of modes, reported as the pressure field $\hat p$, collected at the arrows indicated in panel (a).}
 \label{Fig:NTP}
\end{figure}

Next we define the deformation path as
$$
g=a/2 - a/5 \, \sin(\Psi), \quad \varphi =54^\circ + 20^\circ \cos(\Psi).
$$
This encloses the singular $p11g$-manifold and it leads to the gapped bulk dispersion spectrum shown in the inset of fig.~\ref{Fig:TP}(a). According to our prediction, this loop supports a Chern number $+1$ and, as such, the bulk-boundary correspondence principle \cite{ProdanSpringer2016} assures us of the emergence of topological edge states in a finite geometry, which display a topological spectral flow with the pumping parameter $\Psi$. This is confirmed in the COMSOL simulations reported in fig.~\ref{Fig:TP}(a), where two chiral edge bands located at the opposite ends of finite crystal are clearly visible. Furthermore, samples of the modes as well as a computation of their center of mass confirm their expected localization.
In contrast, if the pumping cycle does not encircle the singular $p11g$-manifold, as fig. \mbox{\ref{Fig:NTP}} exemplifies, then the topological edge state inside the band gap always remains localized on one of the sides of the finite chain of resonators and never crosses the band gap.

In conclusion, we announced a {guiding} principle that enables one to {identify adiabatic cycles with high potential for Thouless pumping,} without making appeal to any analytic tight-binding model. Indeed, the {topological} cycles described in Figs.~\ref{Fig:AC1} and \ref{Fig:AC2} were produced using only geometric considerations. Shunting the need of analytic calculation can have great practical implications. For example, our extremely simple topological adiabatic cycle for the C-shaped resonators (see \cite{Suppl} for an animation) would have been hard to guess from an analytic model, yet we discover it without much effort using the new geometric principles. 
We anticipate that {our guiding} principle {will be fruitful as well for} wallpaper groups and crystallographic groups in 3-dimensions, {and even for} finite highly symmetric molecules. The principle may be {also} relevant to the electron-phonon coupling in quantum materials. For example, deformations of the crystalline or molecular structures that encircle manifolds of enhanced symmetries may supply mechanisms for quantized charge transfers across extended systems.

\acknowledgments
EP acknowledges financial support from US National Science Foundation through the grants CMMI-2131760 and DMR-1823800, and US Army Research Office through contract W911NF2310127. Support from the EIPHI Graduate School [Contract No. ANR-17-EURE-0002] and from ANR PNanoBot [Contract No. ANR-21-CE33-0015] is also acknowledged.

\bigskip
COMSOL simulations are available from the authors upon reasonable request.

\end{document}